\documentclass[a4paper]{jpconf}
\usepackage{graphicx}
\usepackage[square,sort,comma,numbers]{natbib}
\usepackage{caption}
\usepackage{subcaption}
\usepackage{longtable}
\usepackage{float}
\usepackage{wrapfig}
\usepackage{latexsym}
\usepackage{hyperref}
\usepackage{amssymb,amsmath}
\usepackage[alsoload=hep]{siunitx}
\usepackage[nolist,nohyperlinks]{acronym}
\usepackage{xspace}
\newcommand{\genfit}{\textsc{Genfit}\xspace}

\newcommand{\GEANT}{\textsc{Geant\,3}\xspace}

\newcommand{\ROOT}{\textsc{Root}\xspace}

\def\panda{\leavevmode{\sc\setbox0\hbox{p}\hbox to
    \wd0{\vbox{\hrule\vskip1.2pt\box0}}anda}\xspace}
\def\PANDA{\leavevmode{\sc\setbox0\hbox{p}\hbox to
    \wd0{\vbox{\hrule\vskip1.2pt\box0}}anda}\xspace}

\graphicspath{ {plots/} }

\begin{document}
\title{GENFIT -- a Generic Track-Fitting Toolkit}

\author{Johannes Rauch$^a$ and Tobias Schl\"uter$^b$}

\address{
\begin{tabular}{ll}
$^a$&Technische Universit\"at M\"unchen, Physik-Department E18\\
&James-Franck-Stra{\ss}e, 85748 Garching, GERMANY\\
$^b$&Ludwig-Maximilians-Universit\"at M\"unchen, Excellence Cluster Universe\\
&Boltzmannstr.\ 2, 85748 Garching, GERMANY
\end{tabular}
}

\ead{j.rauch@tum.de, tobias.schlueter@physik.uni-muenchen.de}

\begin{abstract}
\genfit is an experiment-independent  track-fitting toolkit
that combines fitting algorithms, track representations,
and measurement geometries into a modular framework.
We report on a significantly improved version of \genfit, based on experience gained in the Belle~II, \PANDA, and FOPI experiments.
Improvements concern the implementation of additional track-fitting algorithms, 
enhanced implementations of Kalman fitters, enhanced visualization capabilities, and
additional implementations of measurement types suited for various kinds of tracking detectors. 
The data model has been revised, allowing for efficient track merging, smoothing, residual calculation, alignment, and storage.
\end{abstract}



\renewcommand{\bflabel}[1]{\normalfont{\normalsize{\textbf{#1}}}\hfill}
\newcommand{\RAVE}{\textsc{Rave}\xspace}

\begin{acronym}[RKTrackRep~~~]

  \acro{AVF}{Adaptive Vertex Fitter}
  \acro{AVR}{Adaptive Vertex Reconstructor}

  \acro{basf2}{Belle~II analysis framework}

  \acro{DAF}{deterministic annealing filter}

  \acro{FOPI}{A $4\pi$ detector located at heavy ion research center, GSI}

  \acro{GBL}{general broken lines}

  \acro{MC}{Monte Carlo}
  
  \acro{NDF}{number of degrees of freedom}
  
  \acro{PANDA}[\leavevmode{\sc\setbox0\hbox{p}\hbox to
    \wd0{\vbox{\hrule\vskip1.2pt\box0}}anda}]{\textbf{Anti-P}roton
    \textbf{An}niliation at \textbf{Da}rmstadt }
  \acro{POCA}{point of closest approach}

  \acro{RAVE}[\RAVE]{Reconstruction (of vertices) in Abstract, Versatile Environments}
  \acro{RKTrackRep}[{{\tt RKTrackRep}}]{Runge-Kutta track representation}%
  \acrodefplural{RKTrackRep}[{{\tt RKTrackReps}}]{Runge-Kutta track representations}%

  \acro{TPC}{Time Projection Chamber}
  \acro{TrackCand}[{{\tt TrackCand}}]{track candidate}
  \acro{TrackRep}{track representation}
  \acro{TUM}{Technische Universit\"at M\"unchen}
  
  \acro{UML}{Unified Modeling Language}
    
\end{acronym}

\section{Introduction}
\label{sec-case}

\label{sec-1-1} 
\genfit provides an extensible modular open-source framework that 
performs track fitting and other related tasks and thus eliminates the redundancy of writing track-fitting programs for every experiment \cite{Hoeppner:2009af}.
Smaller experiments, which do not have the manpower to develop their own track fitters, 
and new experiments, which need working tools to  research and develop, are especially encouraged to use \genfit.
\genfit can also be a valuable teaching aid; the 3D display 
excellently  illustrates track fitting.

\label{sec-1-2} 
All particle physics experiments need to identify
and classify  processes based on detector signals.
Combining these signals to recover  particle trajectories is the task
called {\em tracking}:  Suitable collections of measurements
must be combined into track candidates.  
Tracks must be fitted and 
points of common origin or exit, vertices, must be found and fitted.  
Track finding and track fitting are not independent from each other.
Steps of different levels of track finding and refining alternate with track-fitting steps.
Fitted tracks from different tracking subdetectors can be matched and combined into larger tracks;
and single measurements can be appended to existing tracks, which might come from another
subdetector.
Provided suitable collections of measurements, all of these tasks can be performed with \genfit.

\genfit was successfully used during a test of the Belle~II high-level trigger architecture 
and the combined Belle~II vertex-detector readout architecture~\cite{Bilka:2014lla}.  
There it served for online track reconstruction in the data-reduction stage of the high-level trigger, as well as for offline analysis.  
It was also used as the track fitter supplying input to the Millipede~II alignment software \cite{Blobel:2006yh}.

\section{The three pillars of track fitting}
\label{sec-trinity}
Track fitting in \genfit is based on three pillars:
Measurements, track representations, and fitting algorithms.
Measurements serve as objects containing measured coordinates from 
a detector. They provide functions to construct a (virtual) detector plane and to 
provide measurement coordinates and covariance in that plane.
The abstract base class  {\tt AbsMeasurement} defines the interface. \genfit comes with predefined
measurement classes for various detector types, including planar detectors,
drift chambers, and time projection chambers.

For planar detectors, the detector plane is given by the detector geometry,
whereas for wire  and spacepoint measurements, so-called virtual detector planes
are constructed.
Further information can be used here,
allowing one to compensate for detector deformations like plane bending, wire sag,
and misalignment. 
Drift-time corrections and cluster fits (track dependent clustering) are possible.

Track representations combine track-parameterization and track-extrapolation code.
\genfit implements a track representation based on a Runge-Kutta
extrapolator ({\tt RKTrackRep}, see below).

\label{sec:FittingAlgos}
Fitting algorithms use the measurements and track representations to calculate fit results,
which are stored in corresponding objects in the {\tt Track} object.

\section{The track data structure}
\label{sec-2}

\begin{figure}
  \centering
  \includegraphics[width=1.\textwidth]{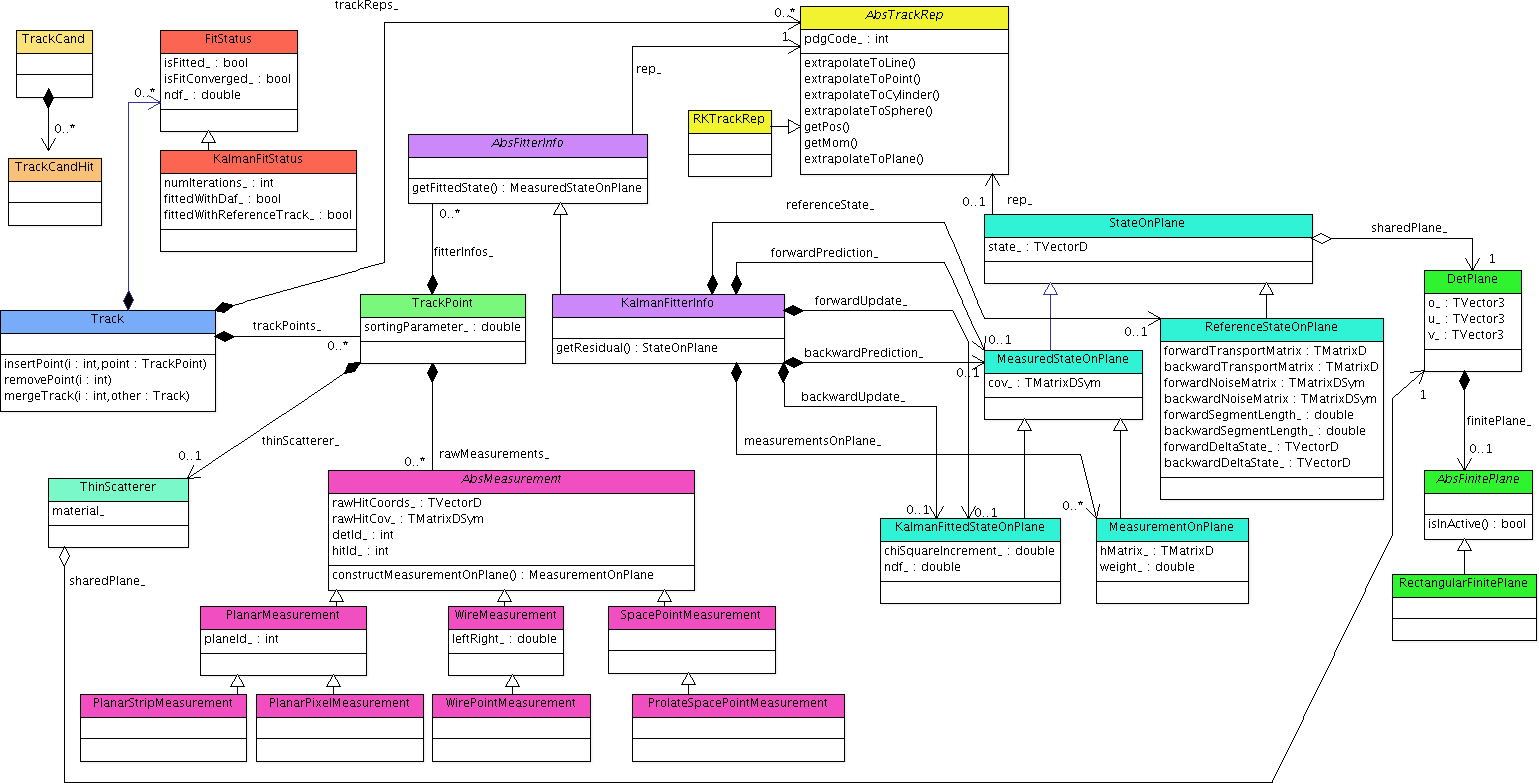}
  \caption{\genfit Unified Modeling Language class diagram.}
  \label{fig:class-diagram}
\end{figure}

All per-track data is kept in the {\tt Track} object (Fig.~\ref{fig:class-diagram}).  
It holds a sequence of {\tt TrackPoint} objects, which can contain measurements, {\tt FitterInfo} objects,  which hold all fitter-specific information, 
and thin scatterers (currently only used in the GBL fitter).

A {\tt Track} contains one or more track representations, representing the particle hypotheses that should be fitted. 
One of them must be selected as the cardinal track representation.
This can  be done by the user or by \genfit, which selects the track representation that best fits the measurements (i.e.\ has the lowest $\chi^2$).
  
The {\tt Track} also contains a {\tt FitStatus} object, which stores general information (number of iterations, convergence, etc.) 
and fit properties ($\chi^2$, \acs{NDF}, $p$~value, track length, etc.).

\label{sec-2-1}
The \ac{TrackCand} serves as a helper class, basically storing indices of raw detector hits
 in  {\tt TrackCandHit} objects, which can also be overloaded by the user to store additional information.
{\tt WireTrackCandHit} objects can store the left-right ambiguity.
A use case is pattern recognition: algorithms can work on raw objects that are lighter weight
than a {\tt Measurement} or implement other features needed by the pattern recognition but not necessarily by \genfit.
\genfit's {\tt MeasurementFactory} can build a {\tt Track} from a {\tt TrackCand}.
This {\tt Track} can then be processed by the various fitting algorithms.

\label{sec-pruning}
After fitting, {\tt Track} objects contain a lot of data: track representations,  {\tt TrackPoint} objects with measurements,  {\tt FitterInfo} objects, etc. 
Usually, not all of this information needs to be stored on disk.
E.g.\ for physics analysis only parameters near the vertex and information on the track quality are necessary, 
whereas alignment needs residual information at all measurements.  
The user can decide which data to keep; e.g.\ only the fitted state of the first {\tt TrackPoint}, and only for the cardinal track representation.

\section{Runge-Kutta track representation}
\label{sec-rktrackrep}

The \ac{RKTrackRep}  is based on a Runge-Kutta extrapolator from \GEANT~\cite{GEANT03}.
An abstract interface class interacts with the detector geometry. Implementations using
\ROOT's {\tt TGeoManager} \cite{root} and {\sc Geant4}'s {\tt G4Navigator} \cite{GEANT04} are available. During fitting, material properties are used to calculate the following effects:
energy loss and energy-loss straggling for charged particles according to the Bethe Bloch formula (code ported from \GEANT);
multiple scattering (according to Ref.\ \cite{Fontana07pn}  or using the Highland formula), where the full noise matrix is calculated;
and soft Bremsstrahlung energy loss and energy-loss straggling for $e^-$ and $e^+$ (code ported from \GEANT).

The step sizes used for the Runge-Kutta extrapolation should be as large as possible to save unnecessary computation,
while still being small enough to keep errors reasonably small.
An adaptive step-size calculation is done in the \ac{RKTrackRep}, 
taking field inhomogeneities and curvature into account.
To calculate material effects correctly, extrapolation stops at material boundaries
and steps can only be so large that a maximum relative momentum loss in the material is not exceeded.

\label{sec:VirtualPlaneConstruction}
\ac{RKTrackRep} provides different methods to find the \ac{POCA} of the track to nonplanar measurements. 
These are used to construct \textit{virtual detector planes}:
\begin{itemize}
  \item \textit{Extrapolate to point} finds the \ac{POCA} of the track to a given spacepoint.
        The virtual detector plane  contains the spacepoint and the \ac{POCA} and is perpendicular to the track.
	
	A \textit{weight matrix} can be used as a metric, defining the space in which the \ac{POCA} 
	will be calculated. By default, the inverted 3D covariance of a spacepoint measurement is used as a metric, 
	which gives correct fitting results also for spacepoints with arbitrary covariance shapes.
	Again, the virtual detector plane  contains the spacepoint and the \ac{POCA} and is perpendicular to the track in the space defined by the metric.

  \item \textit{Extrapolate to line} finds the \ac{POCA} of the track to a given line or wire. 
	The virtual detector plane contains the line and the \ac{POCA}.
	This routine is used for fitting wire measurements.
\end{itemize}
The intersection of the virtual detector plane and the measurement covariance gives the covariance in the plane.

\section{Fitter implementation details}
\label{sec-fitters}

Four different track-fitting algorithms are currently implemented:
two smoothing Kalman filters, one which linearizes the transport around the
state predictions and one which linearizes around a reference track;
a \ac{DAF}; and a \ac{GBL} fitter.

\genfit provides the possibility to store several measurements of the same type in one {\tt TrackPoint},
mainly for using the \ac{DAF} to assign weights to them. 
Wire measurements also produce  two {\tt MeasurementOnPlane} objects,  representing the passage of the particle on either side of the wire. 
These tracks can also be fitted with the Kalman fitters. \genfit provides several ways to handle multiple measurements:
\begin{itemize}
 \item The weighted average of the individual measurements is calculated. This is used by the \ac{DAF}.
  \item The measurement  closest to the state prediction or reference state is selected.
  \item If the {\tt TrackPoint} has a wire measurement, the side that is closest to the prediction or reference is select; otherwise the average is selected.
\end{itemize}
For wire measurements, it turned out that selecting the side closest to the state prediction is the best option for the Kalman fitters.

As convergence criteria, a minimum and maximum number of iterations can be set, which are 2 and 4 by default.
After the minimum number of iterations, \genfit checks if the $p$~value has changed less than a certain amount
since the previous iteration; the default is \num{e-3}.
However, tracks with a $p$~value close to zero are often considered as converged with this criterion, even though the $\chi^2$, albeit big,
is still changing significantly, indicating that the fit is still improving.
This occurs often for tracks that are given bad start parameters.
To cure this issue, a nonconvergence criterion has been introduced: If the relative change in  $\chi^2$ from one iteration to
the next is larger than \SI{20}{\percent}, the fit will continue. Again, this number is user-adjustable.

\subsection{Kalman fitter}
\label{sec-simpleKalman}
The Kalman fitter is adapted from the formulas given in Ref.~\cite{Fruehwirth:2000},
with linearization around the state predictions.
Optionally, a square-root formalism, adapted
from Ref.~\cite{Anderson:1979} to the usecase relevant to
track fitting,  provides greater numerical stability at the
expense of execution time.

\subsection{Kalman fitter with reference track}
\label{sec-refKalman}
State predictions may  stray very far from actual trajectories.
Consequently, linearizing around them is not optimal; material and magnetic field lookups
are also not done at the proper places.
It is therefore common to linearize around reference states \cite{Fruehwirth:2000},
which are calculated by extrapolating the start parameters to all {\tt TrackPoint} objects.  At later iterations, the smoothed states from the
previous iterations are used as linearization points.
However, if the change would be very small, reference states are not updated, saving computing time.
It is also possible to let the fitter sort the measurements along the reference track, which can improve fitting
accuracy.

In addition to the convergence criteria detailed above, 
the fit is regarded as converged if none of the reference states has been updated since the previous iteration.

\subsection{Deterministic annealing filter}
\label{sec-DAF}
The deterministic annealing filter \cite{Fruehwirth1999197} is a powerful tool for the
rejection of outlying measurements. It is a Kalman filter that
uses a weighting procedure between iterations based on the measurement residuals to
determine the proper weights.
The user can select which of \genfit's two Kalman-fitter implementations should be used and specify the annealing scheme.

The \ac{DAF} is also perfectly suited to resolving the left-right ambiguities of wire measurements.
However, a problem can occur:
The weights of the {\tt MeasurementOnPlane} objects  must be initialized. The basic solution is to initialize both left and right measurements with a weight of 0.5.
Effectively the wire positions are taken as measurements in the first iteration, and their covariance
is twice the mean of the individual covariances. 
So all the wire positions have the same covariance, no matter how far from the actual trajectory they are.
This systematically false estimate of the covariances biases the fit.
\genfit implements a novel technique to initialize the weights that improves the fitting efficiency:
Measurements with  larger drift radii  are assigned smaller weights, leading to larger covariances
since the wire position is expected to be farther away from the trajectory.
In contrast, measurements with smaller drift radii, which are closer to the trajectory, get larger weights.

After the annealing scheme, convergence is checked: If the absolute change of all weights 
is less than \num{e-3} (user configurable), the fit is regarded as converged.
Otherwise another iteration with the last temperature is done, until the fit converges or a maximum number of iterations is reached.

\subsection{Generalized broken lines fitter}
\label{sec-GBL}
The generalized broken lines method of track fitting~\cite{Kleinwort:2012np} was implemented especially for the purpose of alignment with Millipede~II.  
It is mathematically equivalent to the Kalman fitter (with thin scatterers instead of continuous materials), 
but fits tracks in their entirety in one step, providing a natural interface to the Millipede~II software.  
For alignment purposes, \genfit also provides a set of interfaces 
for alignment parameters and derivatives which can be implemented by the detector classes.

\section{Vertex reconstruction with GFRave}
GFRave, an interface to the vertex-fitting framework \acs{RAVE}\footnote{\acl{RAVE}},
has been implemented.
\acs{RAVE}~\cite{Waltenberger:2011zz} is a detector-independent toolkit for vertex reconstruction
originally developed for the CMS experiment~\cite{ref:CMS}. 
GFRave takes full advantage of the \genfit material model
and the sophisticated algorithms of \acs{RAVE}, allowing for precise and fast
vertex reconstruction.

\section{Performance}
\label{sec-5}
The execution time of the {\tt GenFitter} module in the Belle~II software framework~\cite{Moll:2011zz}
was measured on a \SI{3.4}{\giga\hertz} office PC in single-threaded operation. 
All code was compiled with {\tt -O3} optimization settings.
Besides fitting all tracks in an event, this module builds data structures for Belle~II's data storage format, incurring an overhead of slightly less than \SI{1}{\ms}.
Single track events were generated with a particle gun, with $\theta = \SI{100}{\degree}$ and a momentum of \SI{0.9}{\GeV} in a constant magnetic field. 
The resulting {\tt Track} objects found by MC-based (perfect) track finding contain $72 \pm 2$ {\tt TrackPoint} objects.
The Kalman fitters have been configured to do 3 to 10 iterations with default convergence criteria, while the \ac{DAF} uses its
default annealing scheme with 5 temperatures.

From the results shown in Tab.\ \ref{tab:performance}, one can see that the Kalman fitter with reference track needs fewer iterations to converge.
Material lookup requires approximately \SI{2.2}{\milli\second} per iteration for the Kalman fitter. The Kalman fitter with reference track is also faster here,
since the reference states are not recalculated if they are close to the smoothed states of the previous iteration.
This is also the reason why the reference Kalman can often finish after 2 iterations.

\begin{table}
\caption{\label{tab:performance}Execution time of the GenFitter module in the Belle~II software framework.}
\begin{center}
\begin{tabular}{llll}
\br 
Fitter & w/o matFX & w/ matFX & $\varnothing$ iterations\\
\mr
Kalman    & \SI{3.4}{\ms} & \SI{10}{\ms} & 3 \\
Reference Kalman & \SI{4.0}{\ms} & \SI{8.2}{\ms} & 2.13 \\
DAF              & \SI{9.4}{\ms} & \SI{17}{\ms} & 6 \\
\br
\end{tabular}
\end{center}
\end{table}

\section{Visualisation}
\label{sec:visualisation}
\genfit features a sophisticated 3D event display, which allows one to visualize fitted tracks.
Detector geometry, measurements, detector planes, reference tracks, forward and backward fits (predictions and updates), smoothed tracks,
and covariances of measurements and tracks can be drawn.
Tracks can be refitted with different algorithms and settings, and fit results can be viewed
instantly.

\begin{figure}
        \centering
        \begin{subfigure}[t]{0.475\textwidth}
                \includegraphics[trim =  16cm 2.25cm 20cm 4.6cm, clip, width=\textwidth]{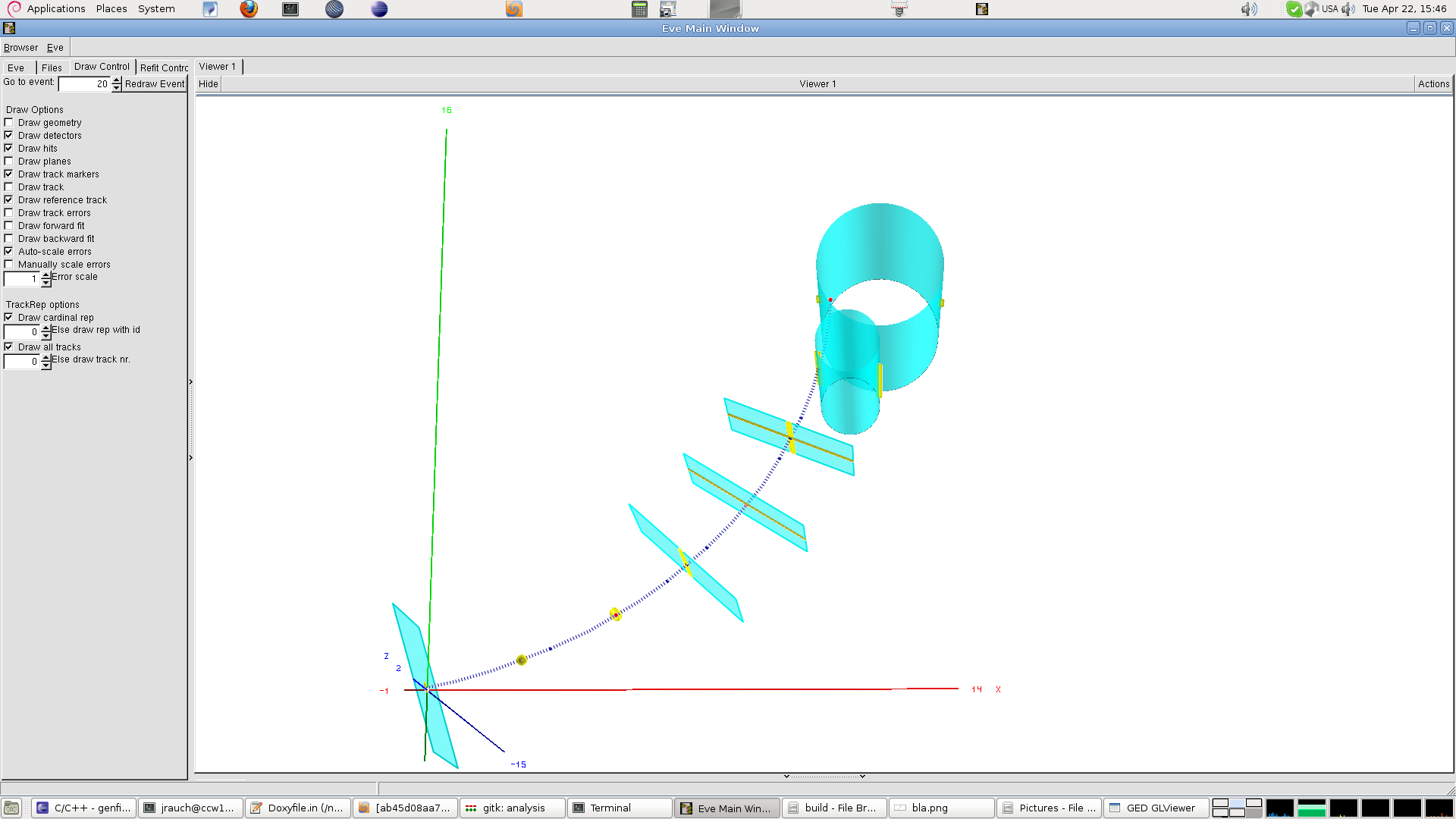}
                \caption{Measurements with covariance (yellow), planar detectors and drift isochrones (cyan), and reference track (blue).}
        \end{subfigure}%
	\quad
        \begin{subfigure}[t]{0.475\textwidth}
                \includegraphics[trim =  16cm 2.25cm 20cm 4.6cm, clip, width=\textwidth]{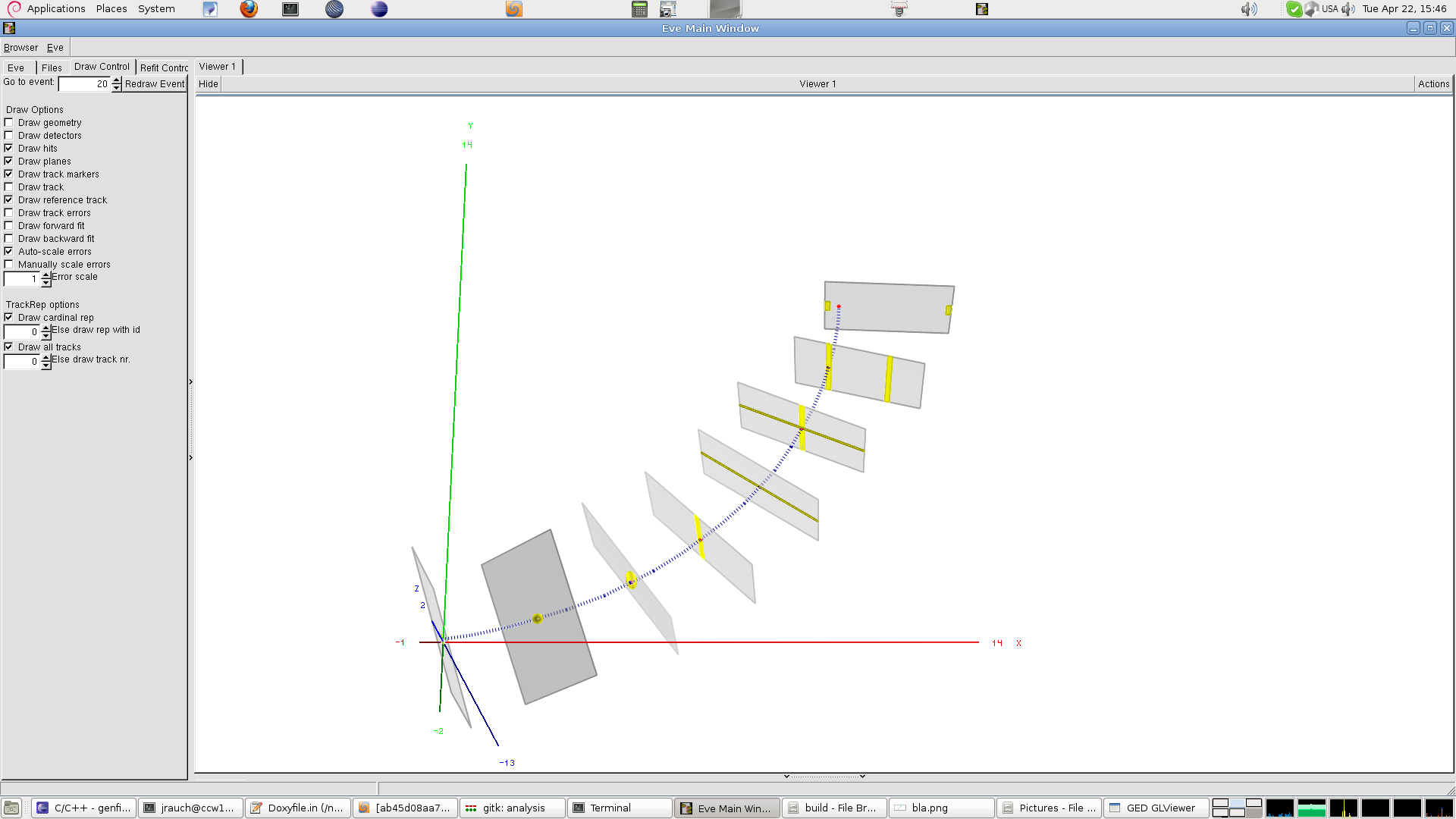}
                \caption{Detector planes (gray). For the spacepoint and wire measurements virtual detector planes have been constructed.}
        \end{subfigure}

        \begin{subfigure}[t]{0.475\textwidth}
                \includegraphics[trim =  16cm 2.25cm 20cm 4.6cm, clip, width=\textwidth]{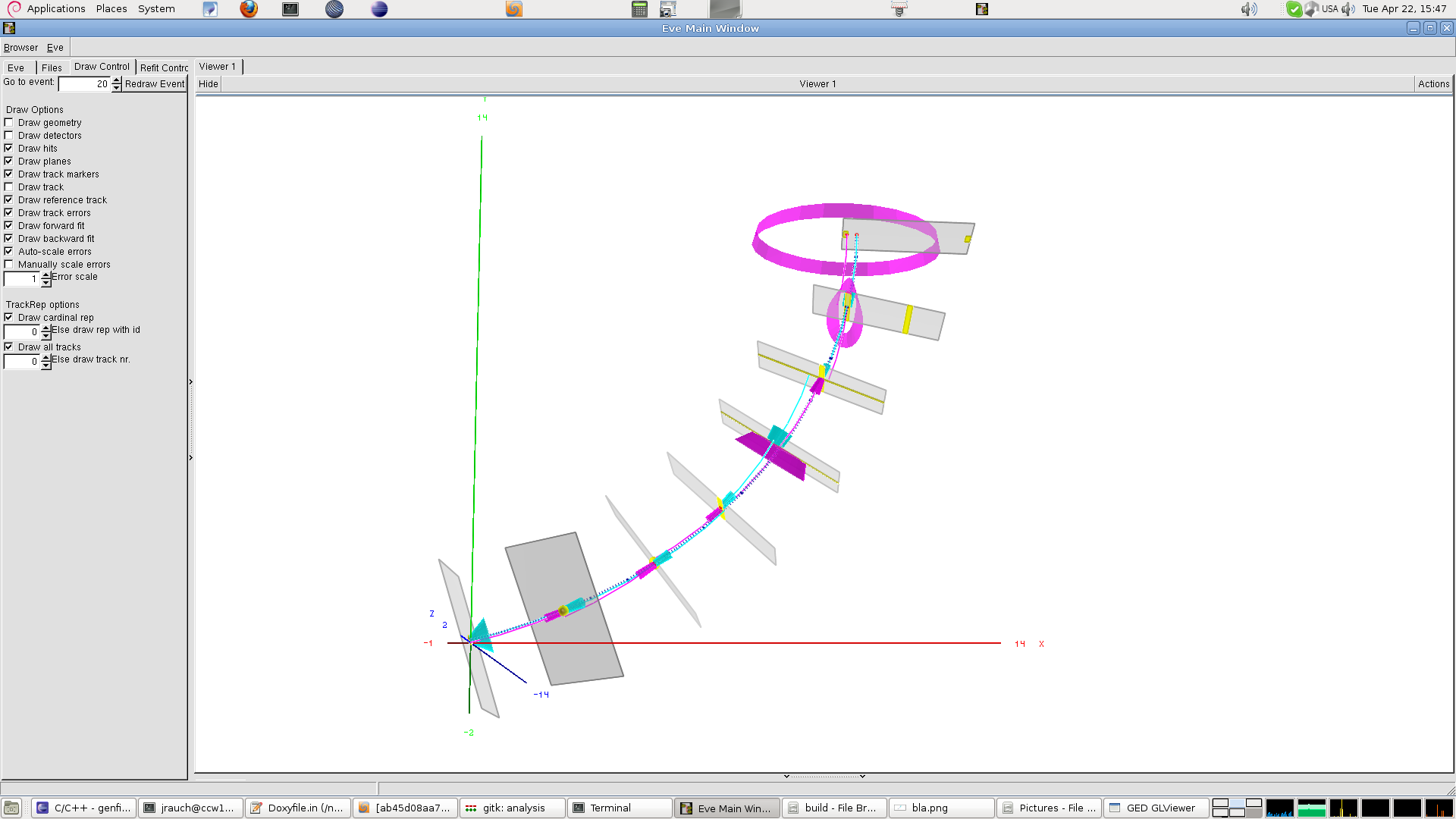}
                \caption{Forward (cyan) and backward (magenta) fit with covariances of the state updates.}
        \end{subfigure}
	\quad
        \begin{subfigure}[t]{0.475\textwidth}
                \includegraphics[trim =  16cm 2.25cm 20cm 4.6cm, clip, width=\textwidth]{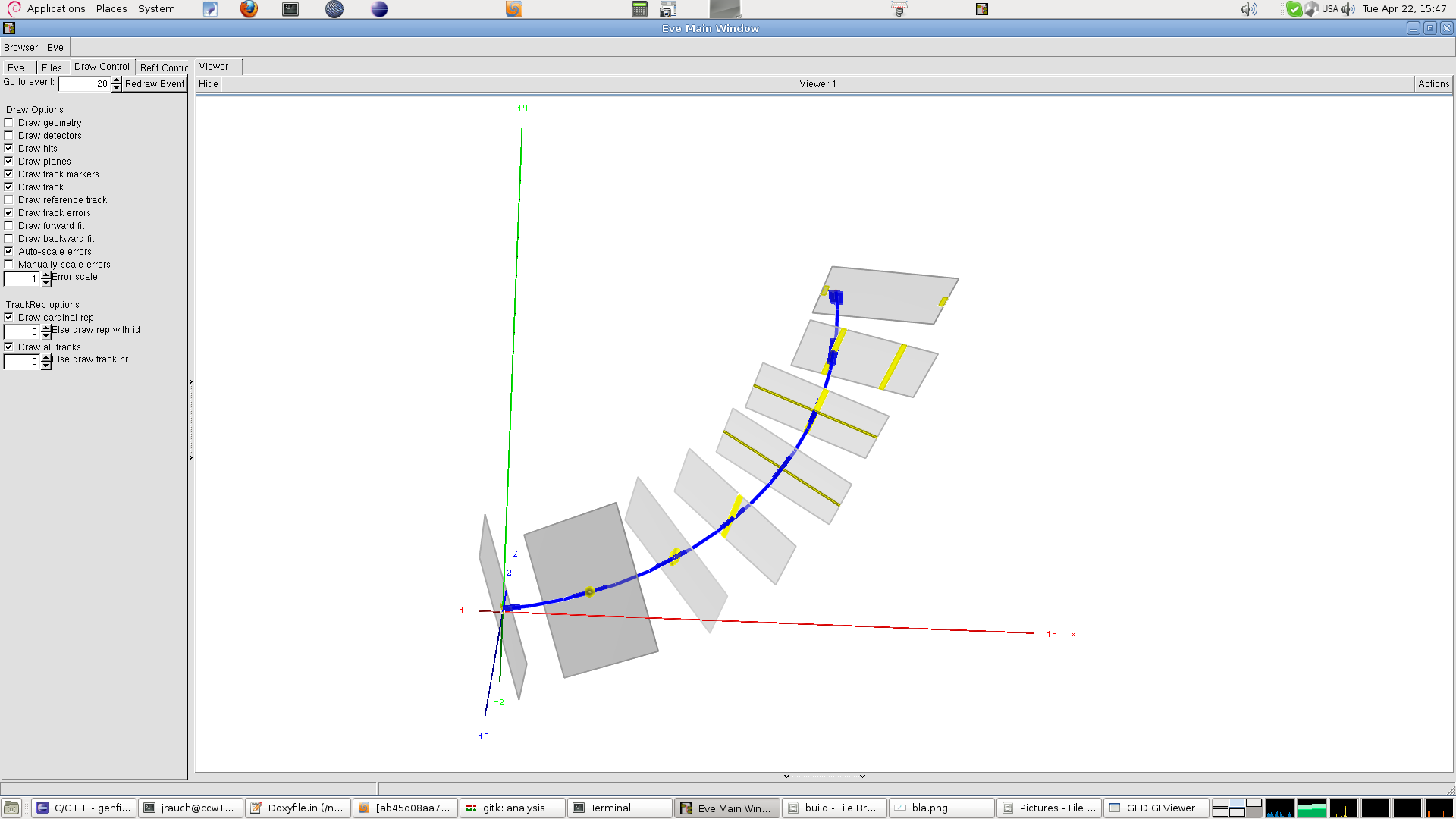}
                \caption{Smoothed track with covariance (blue).}
        \end{subfigure}
        \caption{\genfit event display screenshots.}\label{fig:eventDisplay}
\end{figure}

Fig.\ \ref{fig:eventDisplay} shows the fit of a set of measurements with the Kalman fitter with reference track.
For demonstration purposes, the different measurement types supported by \genfit are used (starting from the origin): planar pixel measurement,
 spacepoint measurement, prolate spacepoint measurement, two perpendicular planar strip measurements, double-sided planar strip measurement, 
 wire measurement, and wire measurement with second coordinate measurement.

\section*{Acknowledgments}
This research was supported by the DFG cluster of excellence ``Origin and Structure of the Universe.''
T.S.\ was supported under BMBF Contract 05H12WM8.

\bibliographystyle{plainurl}
\bibliography{genfit}

\end{document}